\documentclass[12pt,a4paper,twoside]{article}
\usepackage{epsfig}
\usepackage{baltlat1}
\usepackage{wrapfig}
\begin{document}
\ \
\vspace{0.5mm}

\setcounter{page}{1}
\vspace{8mm}

\titlehead{Baltic Astronomy, vol.12, XXX--XXX, 2003.}

\titleb{SPECTRAL-AGEING ANALYSIS OF SELECTED\\ DISTANT GIANT RADIO 
GALAXIES}

\begin{authorl}
\authorb{M.~Jamrozy}{1} and
\authorb{J.~Machalski}{2}

\end{authorl}

\begin{addressl}
\addressb{1}{Radio Astronomical Institute, Bonn University, 
Auf dem H\"ugel 71,\\ 53-121 Bonn, Germany }

\addressb{2}{Astronomical Observatory, Jagellonian University,
ul. Orla 171,\\ 30-244 Cracow, Poland}

\end{addressl}

\submitb{Received: xx October 2003}

\begin{abstract}
Very deep 4.9~GHz observations with the VLA are used to
supplement available radio maps at the frequencies of 325~MHz (WENSS 
survey) and 1.4~GHz (NVSS survey) in order to study the synchrotron 
spectra and radiative ages of relativistic particles in the extended 
lobes of distant {\sl giant} radio galaxies selected from the sample 
of Machalski et al. (2001). 
\end{abstract}


\resthead{Spectral-ageing of Giant Radio Galaxies}
{M.~Jamrozy and J.~Machalski}



\sectionb{1}{INTRODUCTION AND THE SAMPLE}

Our analysis of spectral ages of radiating particles in the opposite 
lobes of faint {\sl giant} radio galaxies follows the classical 
spectral-ageing method used to estimate radiative ages and expansion 
speeds in a large set of powerful 3CR sources (Alexander \& Leahy 1987, 
{\footnotesize MNRAS, 225, 1}; Leahy et al. 1989, {\footnotesize MNRAS, 
239, 401}; Carilli et al. 1991, {\footnotesize ApJ, 383, 554}; Liu et 
al. 1992, {\footnotesize MNRAS, 257, 545}, as well as in samples of 
low- and medium-luminosity radio galaxies (e.g. Klein et al. 1995, 
{\footnotesize A\&A, 303, 427}; Parma et al. 1999, 
{\footnotesize A\&A, 344, 7}). Those sources were used by Machalski et 
al. (2003) as the observational base for a statistical study of the 
dynamical evolution of the population of FRII-type radio sources 
(cf. also Machalski et al.; these proceedings).

Three sources from the sample of Machalski et al. (2001) were selected 
for the present spectral analysis; their basic data are given in 
columns 2, 3, 4 and 5 of Table~1. The spectral analysis of the radio 
lobes of the above sources has based on the radio maps at 325 and 
1400~MHz available from the surveys: WENSS (Rengelink et al. 1997) and 
NVSS (Condon et al. 1998). For the low-frequency extent of the radio 
spectra the flux  density data provided in the 7C catalogues (Riley et 
al. 1999, {\footnotesize MNRAS, 306, 31}) were used. There was a lack 
of relevant data at frequencies higher than 1.4~GHz, so we aimed to 
observe the above {\sl giant} sources at 5 GHz with the VLA in its 
D-array.

\sectionb{2}{4.9 GHz MAPS AND RADIO SPECTRA OF THE 3 `GIANTS'}

The observations were carried out in February 2003. During the total 
exposure time from 50 to 70~min for each of the two extended radio
lobes, the {\sl rms} noise level varied from about 32 to 20
{\footnotesize$\mu$}\,Jy\,beam{\footnotesize$^{-1}$}. Calibrated 
brightness distribution of both lobes, corrected   for the `primary 
beam' attenuation and convolved to the common angular resolution of 
20''$\times$20'', has been combined into the final maps. The maps are 
shown in Fig.~1a,\,b,\,c. The sensitivity achieved allowed us to 
detect a weak radio core in J0912+3510, which was crucial to confirm 
the host galaxy identification for this source.

\begin{figure}[h]
\centerline{\psfig{figure=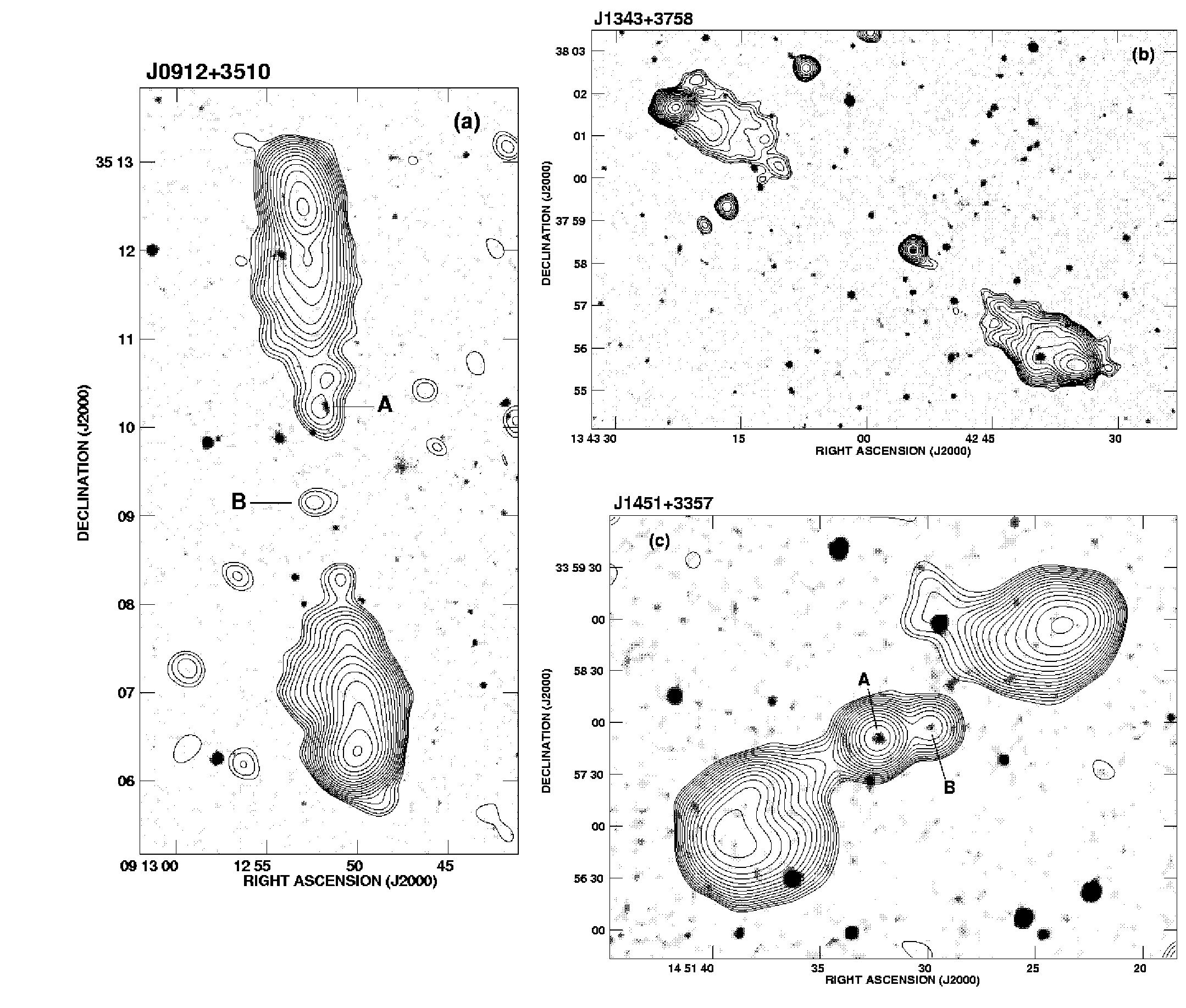,width=100truemm,angle=0,clip=}}
\captionb{1}{{\footnotesize 4.9~GHz VLA contour maps with the FWHM of 
20$^{\prime\prime}$$\times$20$^{\prime\prime}$\,overlayed on the 
optical DSS fields: {\bf a)} for J0912+3510, {\bf b)} for J1343+3758, 
and {\bf c)} for J1451+3357. The identified host galaxies are marked 
with `A'. The faint compact components unrelated to the investigated 
sources are marked with `B'}}
\end{figure}

\sectionb{3}{SPECTRAL AGEING ANALYSIS}

For the spectral analysis, the maps at all frequencies were convolved 
to the common beamwidth of 60$^{\prime\prime}$. Then the spectrum in 
different parts of the lobes has been determined by integration of the 
total intensity along strips perpendicular to the source axis. The 
strips are one common beam across providing independent data points. 
In each lobe the spectrum steepening is clearly visible.

The time elapsed since the particles were last accelerated is 
calculated assuming the Jaffe \& Perola (1973, {\footnotesize A\&A, 26, 
423}) model. The magnetic field strength in the lobes, 
{\footnotesize$B_{eq}$}, is calculated on the assumption of minimum 
energy using the method outlined by Miley (1980, 
{\footnotesize ARA\&A, 18, 165}). Their total radio luminosity is 
integrated between 10~MHz and 100~GHz. The volume of the lobes is 
calculated assuming a cylindrical geometry with the base diameter 
equal to the average width of the lobes measured on the WENSS and NVSS 
maps (column~5 of Table~1) using the prescription of Leahy \& Williams 
(1984, {\footnotesize MNRAS, 210, 929}).

The results of this analysis are shown in Fig.~2a,\,b,\,c where for 
each lobe of the investigated sources (i) the spectral index between 
325 and 4885~MHz, and (ii) the radiative age have been plotted against
distance measured along the source axis from the hotspot. The best 
fits to the age data points are shown by solid curves. The dashed 
curves mark the best fits to an upper limits of the particle age 
calculated with a substitution of {\footnotesize$B_{MBR}/\sqrt{3}$} 
instead of {\footnotesize$B_{eq}$}, where {\footnotesize$B_{MBR}$} is 
the magnetic field equivalent of the 2.7~K microwave background 
radiation. This corresponds to the phase in which a lobe is no longer 
overpressured in respect to an external pressure of the intergalactic 
medium.

\begin{figure}
\vskip2mm
\centerline{{\psfig{figure=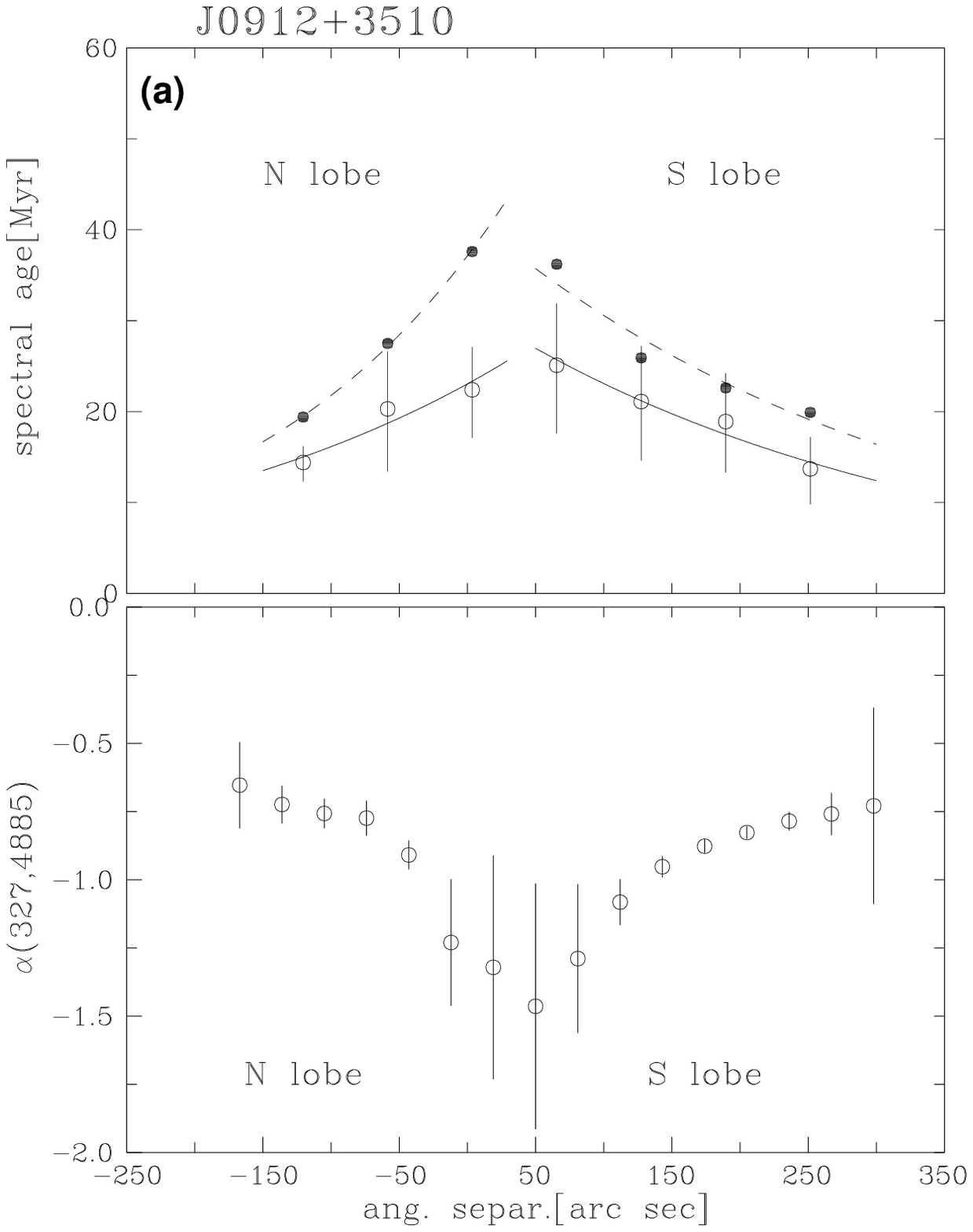,width=45truemm,angle=0,clip=}}
\hskip2mm
{\psfig{figure=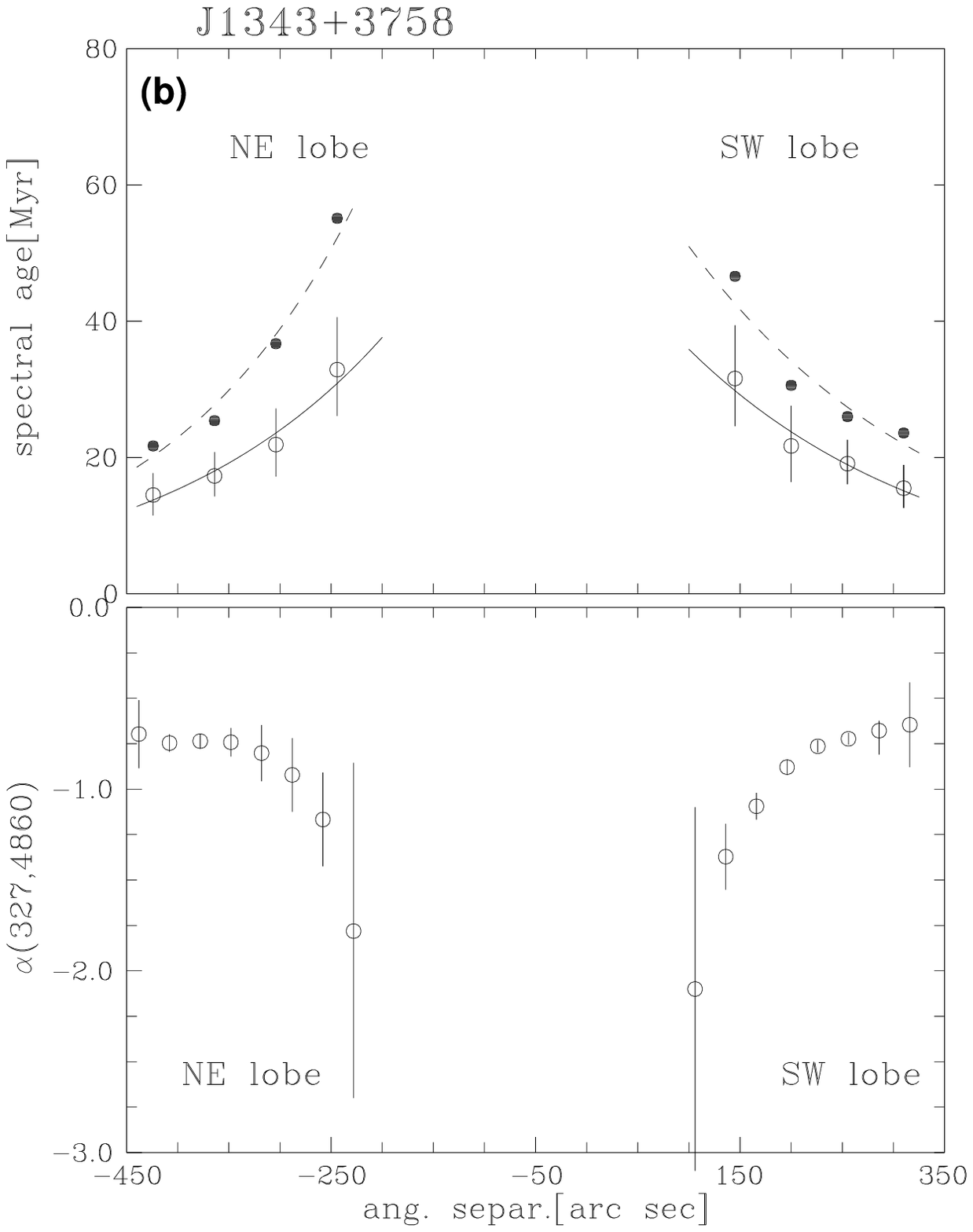,width=45truemm,angle=0,clip=}}
\hskip2mm
{\psfig{figure=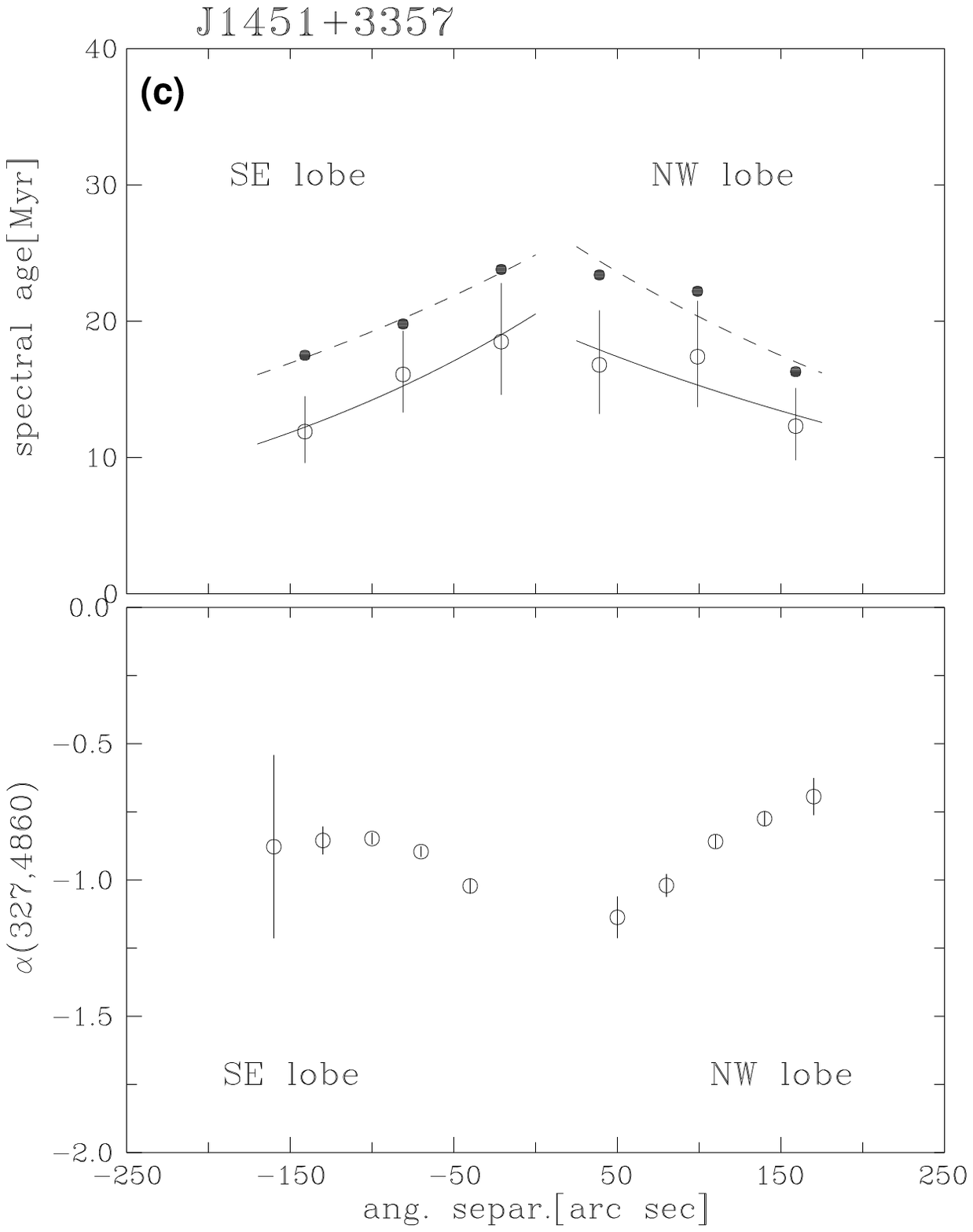,width=45truemm,angle=0,clip=}}}
\captionb{2}{{\footnotesize Spectral index {\footnotesize$\alpha_{325}^{4885}$}
(lower panel), and radiative age (upper panel) along the source axis 
{\bf a)} for J0912+3510, {\bf b)} for J1343+3758, and {\bf c)} for 
J1451+3357. The solid curves show the best fit to the data points, 
while the dashed curves indicate the upper limit of the radiative 
(synchrotron) age [cf. the text]}}
\end{figure}

A comparison of these synchrotron ages of the oldest particles with 
the dynamical age estimates provided by Chy\.{z}y et al. (these 
proceedings) is given in Table~1 (columns 6, 7, 8). The radiative
ages of the oldest particles range from about 15 to 40~Myr, and are 
similar to relevant ages of normal-size radio galaxies at comparable 
redshifts. A comparison of the above ages and the dynamical ages 
derived as above suggests that the mean backflow velocity is 1 $\sim$ 3 
of the mean advance velocity of the jet's head with respect to the 
surrounding intergalactic medium, i.e. the dynamical age of a {\sl 
giant} radio galaxy can be 2 $\sim$ 4 times higher than the 
synchrotron age of the oldest radiating particles.

\begin{center}
\vbox{\footnotesize
{\smallbf\ \ Table 1.}{\small\ The basic data of selected radio 
galaxies calculated for {\footnotesize$H_{0}=50$}\,km\,s{\footnotesize$^{-1}$}\,Mpc{\footnotesize$^{-1}$}
and {\footnotesize$q_{0}=0.5$}, and their radiation ages vs. dynamical 
age estimates, as well as the resultant expansion velocity; 
($^{*}$) -- taken from Machalski et al. (2003)}

\begin{tabular}{llccccccc}
Source   &{\footnotesize$z$} &{\footnotesize$\log L_{1.4}$} 
&{\footnotesize$D$} &{\footnotesize$b$} &
{\footnotesize$t_{rad}$} & {\footnotesize$t_{dyn}$} & 
{\footnotesize$\frac{t_{dyn}}{t_{rad}}$} & {\footnotesize$v_{h}/c$}\\
         &  &[W\,Hz{\footnotesize$^{-1}$}\,sr{\footnotesize$^{-1}$}]& [Mpc] & [kpc] &
[Myr] & [Myr]\\
           &    &    &     &      \\
J0912+3510 & 0.2489 & 24.55 & 1.84 & 372 & 25 & 91 & 3.6 & 0.033\\
J1343+3758 & 0.2267 & 24.42 & 3.14 & 502 & 33 & 94$^{*}$ & 2.8 & 0.054\\
J1451+3357 & 0.3251 & 24.74 & 1.41 & 344 & 19 & 80 & 4.2 & 0.029
\end{tabular}
}
\end{center}

ACKNOWLEDGMENTS.\ 
MJ acknowledges the financial support from EAS.
\goodbreak

References\\
Condon~J.~J., Cotton~W.~D., Greisen~E.~W., et al. 1998, AJ, 115, 1693 (NVSS)\\
Machalski~J., Jamrozy~M., Zola~S. 2001, A\&A, 371, 445\\
Machalski~J., Chy\.{z}y~K.~T., Jamrozy~M. 2003 (submitted for MNRAS, astro-ph/0210546)\\
Rengelink~R., Tang~Y., de Bruyn~A.~G., et al. 1997, A\&AS, 124, 259\\
\end{document}